\documentclass[twocolumn,showpacs,preprintnumbers,superscriptaddress,amsthm,amsmath,amssymb]{revtex4}
\usepackage{graphicx}
\usepackage{dcolumn}
\usepackage{bm}
\usepackage{color}
\usepackage{multirow}
\usepackage{CJK}
\usepackage{ltxtable}
\usepackage{rotating}
\usepackage{array}
\usepackage{booktabs,longtable}
\usepackage[colorlinks,linkcolor=blue,anchorcolor=blue,citecolor=red]{hyperref}
\usepackage{setspace}

\begin{document}

\title{Inner fission barriers of uranium isotopes in the deformed relativistic Hartree-Bogoliubov theory in continuum}

\author{Wei Zhang}
\affiliation{School of Physics and Laboratory of Zhongyuan Light, Zhengzhou University, Zhengzhou 450001, China}

\author{Jin-Ke Huang}
\affiliation{School of Physics and Laboratory of Zhongyuan Light, Zhengzhou University, Zhengzhou 450001, China}

\author{Ting-Ting Sun}
\email{ttsunphy@zzu.edu.cn}
\affiliation{School of Physics and Laboratory of Zhongyuan Light, Zhengzhou University, Zhengzhou 450001, China}

\author{Jing Peng}
\email{jpeng@bnu.edu.cn}
\affiliation{Department of Physics, Beijing Normal University, Beijing 100875, China}

\author{Shuang Quan Zhang}
\affiliation{State Key Laboratory of Nuclear Physics and Technology, School of Physics, Peking University, Beijing 100871, China}

\date{\today}

\begin{abstract}
The inner fission barriers of the even-even uranium isotopes from the proton to the neutron drip line are studied with the deformed relativistic Hartree-Bogoliubov theory in continuum. A periodic evolution for the ground state shapes is shown with the neutron number, i.e., spherical shapes at shell closures $N=$126, 184, 258, and prolate dominated shapes between them. In analogy to the shape evolution,
the inner fission barriers also exhibit a periodic behavior: peaks at the shell closures and valleys in the mid-shells. The triaxial effect to the inner fission barrier is evaluated using the triaxial relativistic mean field calculations plus a simple BCS method for pairing. With the triaxial correction included, good consistency in the inner barrier heights is found with the available empirical data.
Besides, the evolution from the proton to the neutron drip line is in accord with the results by the multi-dimensionally constrained relativistic mean field theory. A flat valley in the fission barrier height is predicted around the neutron-rich nucleus $^{318}$U which may play a role of fission recycling in the astrophysical $r$-process nucleosynthesis.
\end{abstract}



\maketitle


\section{Introduction}
Nuclear fission is one of the most significant topics in nuclear physics, which plays a key role in the studies of superheavy synthesis and the astrophysical $r$-process nucleosynthesis. It has attracted wide attentions by the nuclear theoretical studies as well as experimental investigations in the large-scale facilities around the world. The inner fission barrier is one of the critical quantities, as a 1-MeV variance in height could lead to several orders of magnitude difference in the fission half-life.
However, until now, very few experimental data have been obtained, so that the corresponding theoretical studies are important.

Many theoretical models have been used to study fission barriers, such as the macroscopic-microscopic (MM) model~\cite{HOWARD1980,MYERS1996,Moller2015}, the extended Thomas-Fermi plus Strutinsky integral (ETFSI) method~\cite{MAMDOUH2001}, and the density functional theories in non-relativistic framework~\cite{DELAROCHE2006,Goriely2009,Giuliani2013,Rodriguez2014}
and relativistic framework~\cite{Burvenich2004,Karatzikos2010,Lu2012,Lu2014,Tao2017,Ren2022,Deng2023,RenZX2022PRL,FOP2024}.
Nowadays, the covariant density functional theory~(CDFT) has attracted intense attentions for its inherent merits of Lorentz symmetry~\cite{Ring2012,Meng2021,Meng2016} and achieved great successes in describing a variety of nuclear properties in a microscopic way~\cite{Ring1996,Vretenar2005,Meng2006,Meng2015,Meng2016}, such as nuclear masses and radii~\cite{Geng2005,Pena2016,ZhangKY2022,Guo2024,Wu2024}, half-lives~\cite{Niu2013,Wang2016,Marketin2016}, nuclear magicity~\cite{ZhangW2005,LiuJ2020PLB,ZhangKY2023}, pseudospin symmetry~\cite{Liang2015,Sun2017PRC,Sun2023PLB,Sun2024}, spin symmetry in antinucleon spectrum~\cite{ZhouSG2003PRL}, nuclear rotations~\cite{Zhao2011,MengJ2013FOP,Zhao2015,WangYP2023}, shape coexistence and shape transition~\cite{MengJ2006PRC,Z2018}, level density~\cite{Z2023,JiangXF2024PLB}, low-lying spectrum~\cite{Lizp2012,Yao2014}, and single-particle resonances~\cite{Guo2016,Sun2020PRC,Sun2021NST}, as well as hypernuclear properties~\cite{Lu2011PRC,Lu2014PRC,Sun2018,Sun2021SC}.
Focusing on the fission study, both the static fission properties and the fission dynamics have been investigated by CDFT~\cite{Burvenich2004,Karatzikos2010,Lu2012,Lu2014,Tao2017,Ren2022,Deng2023,RenZX2022PRL,FOP2024}.
For example, by using the multi-dimensionally constrained relativistic mean field (MDC-RMF) theory~\cite{Lu2012,Lu2014}, the fission properties of the even-even uranium isotopes have been studied~\cite{Deng2023}.

In recent years, starting from the CDFT, the deformed relativistic Hartree-Bogoliubov theory in continuum (DRHBc) has been developed, which can treats the deformation, pairing correlations and continuum effects simultaneously~\cite{ZhouSG2010,Lilulu2012}.
Until now, it has been successfully applied to study both stable and weakly-bound nuclei all over the nuclear chart for many interesting topics, such as
the prediction of shape-decoupling phenomenon in deformed halo nuclei~\cite{ZhouSG2010,Lilulu2012,SunXX2020,SunXX2021prc,ZhangKY2023PLB},
the solution of the puzzles concerning the radius and neutron configuration in $^{22}$C~\cite{SunXX2018},
the number of particles in the classically forbidden regions for magnesium isotopes~\cite{ZhangKY2019},
the dependence of the multipole expansion order~\cite{Congpan2019},
the deformation effects on the neutron drip line~\cite{Papakonstantinou2021},
the shape evolution, shape coexistence and prolate-shape dominance~\cite{Kim2022,Choi2022,Mun2023,ZhangXY2023,GuoPeng2023},
the evolution of shell closures~\cite{ZhangKY2023,ZhengRuyou2024},
the stability peninsulas beyond the neutron drip line~\cite{Zhang_2021,CongPan2021,Hexiaotao2021},
the one-proton emission from $^{148-151}$Lu~\cite{Xiaoyang2023},
the optimization of Dirac Woods-Saxon (DWS) basis~\cite{ZhangKY2022prc},
the rotational excitations of exotic nuclei with angular momentum projection~\cite{SunXX2021PRC,SunXX2021science},
the reaction and charge-changing cross sections of light nuclei with the Glauber model~\cite{Jing2022,JWZhao2023,An2024},
and the dynamical correlation with a two-dimensional collective Hamiltonian method~\cite{Sun2022}.
Recently, efforts have been made to construct a DRHBc mass table that takes into account both deformation and continuum effects~\cite{ZhangKY2020,ZhangKY2022,Pan2022,Guo2024}.
However, it is still absent for the application of DRHBc theory in the study of fission barrier.

In this work, shape evolution and the inner fission barrier for even-even uranium isotopes from the proton drip line to the neutron drip line will be studied based on the DRHBc theory. In principle, to extract the height of the fission barrier $B_{\rm f}$, i.e., the energy difference between the global ground state and the respective saddle, one needs to do calculations in the multidimensional deformation space. Besides the most important axial quadrupole deformation $\beta_{2}$, degrees of freedom such as triaxial deformation $\gamma$ and octupole deformation $\beta_{3}$ are also indispensable. For the inner fission barrier, it was found by various models that the inclusion of triaxial deformation could reduce its height up to a few MeVs~\cite{Moller1970,Randrup1976,Girod1983,Rutz1995,Moller2009,Abusara2010,Schunck2014,Zhou2016}.
As the DRHBc theory is limited to the axial symmetry, in this work, to evaluate the triaxial effect,
the triaxial relativistic mean field calculations plus a simple BCS~(RMF+BCS)~method for pairing will also be performed. With such a triaxial correction, the inner fission barrier heights will be compared with the available empirical data and results of FRLDM~\cite{Moller2009}, ETFSI~\cite{MAMDOUH2001}, HFB-14~\cite{Goriely2009} and MDC-RMF~\cite{Deng2023} models.

The paper is organized as the following. Sec.~\ref{section:Theory} briefly introduces the theoretical framework of DRHBc, Sec.~\ref{section:Results} presents the results and discussion, and Sec.~\ref{section:Summary} provides a brief summary and perspective.

\section{Theoretical Framework}
\label{section:Theory}
Detailed formalism of the DRHBc theory can be found in Refs.~\cite{ZhouSG2010,Lilulu2012,ZhangKY2020}. Here a brief introduction is presented. In the DRHBc theory, the relativistic Hartree-Bogoliubov (RHB) equation that treats the mean field and pairing correlations self-consistently reads~\cite{Kucharek1991},
\begin{equation}
	\label{RHB}
	\left(\begin{array}{cc}
		h_D-\lambda & \Delta \\
		-\Delta^* & -h_D^*+\lambda
	\end{array}\right)\left(\begin{array}{c}
		U_k \\
		V_k
	\end{array}\right)=E_k\left(\begin{array}{c}
		U_k \\
		V_k
	\end{array}\right),
\end{equation}
where $E_k$, $h_D$, $\lambda$, and $(U_k, V_k)^T$ are the quasiparticle energy, the Dirac Hamiltonian, the Fermi energy, and the quasiparticle wave function, respectively. The Dirac Hamiltonian $h_D$ is given by
\begin{equation}
	\label{h_d}
	h_D=\boldsymbol{\alpha} \cdot \boldsymbol{p}+\beta(M+S(\boldsymbol{r}))+V(\boldsymbol{r}),
\end{equation}
 where $S(\boldsymbol{r})$ and $V(\boldsymbol{r})$ are the scalar and vector potentials, respectively.
The pairing potential for particle-particle channel reads
\begin{equation}
	\Delta_{k k^{\prime}}\left(\boldsymbol{r}, \boldsymbol{r}^{\prime}\right)=-\sum_{\tilde{k} \tilde{k}^{\prime}} V_{k k^{\prime}, \tilde{k} \tilde{k}^{\prime}}^{p p}\left(\boldsymbol{r}, \boldsymbol{r}^{\prime}\right) \kappa_{\tilde{k} \tilde{k}^{\prime}}\left(\boldsymbol{r}, \boldsymbol{r}^{\prime}\right),
\end{equation}
with the pairing tensor $\kappa=V^* U^T$ and
the density-dependent zero-range pairing interaction
\begin{equation}
	V^{p p}\left(\boldsymbol{r}, \boldsymbol{r}^{\prime}\right)=\frac{V_0}{2}\left(1-P^\sigma\right) \delta\left(\boldsymbol{r}-\boldsymbol{r}^{\prime}\right)\left(1-\frac{\rho(\boldsymbol{r})}{\rho_{\mathrm{sat}}}\right),
\end{equation}
where $\rho_{\mathrm{sat}}$ is the nuclear saturation density and $V_0$ is the pairing strength.

In the DRHBc theory, the potentials and densities are expanded in terms of the Legendre polynomials,
\begin{equation} \label{lmax}
	f(\boldsymbol{r})=\sum_\lambda f_\lambda(r) P_\lambda(\cos \theta), \quad \lambda=0,2,4, \cdots
\end{equation}
Meanwhile, in order to properly consider the continuum effect, especially for nuclei close to drip lines,
the deformed RHB equation~(\ref{RHB}) is solved in a spherical DWS basis~\cite{Zhou2003prc}, the basis wave functions of which have proper asymptotic behaviors in large coordinates. After solving the RHB equation self-consistently, the total binding energy $E_{\rm tot}$, the radius, the intrinsic multipole moments $Q_\lambda$ along with the deformation parameters can be calculated.

\section{Results and Discussion}
\label{section:Results}

\begin{figure*}[t!]
\includegraphics[width=0.8\linewidth]{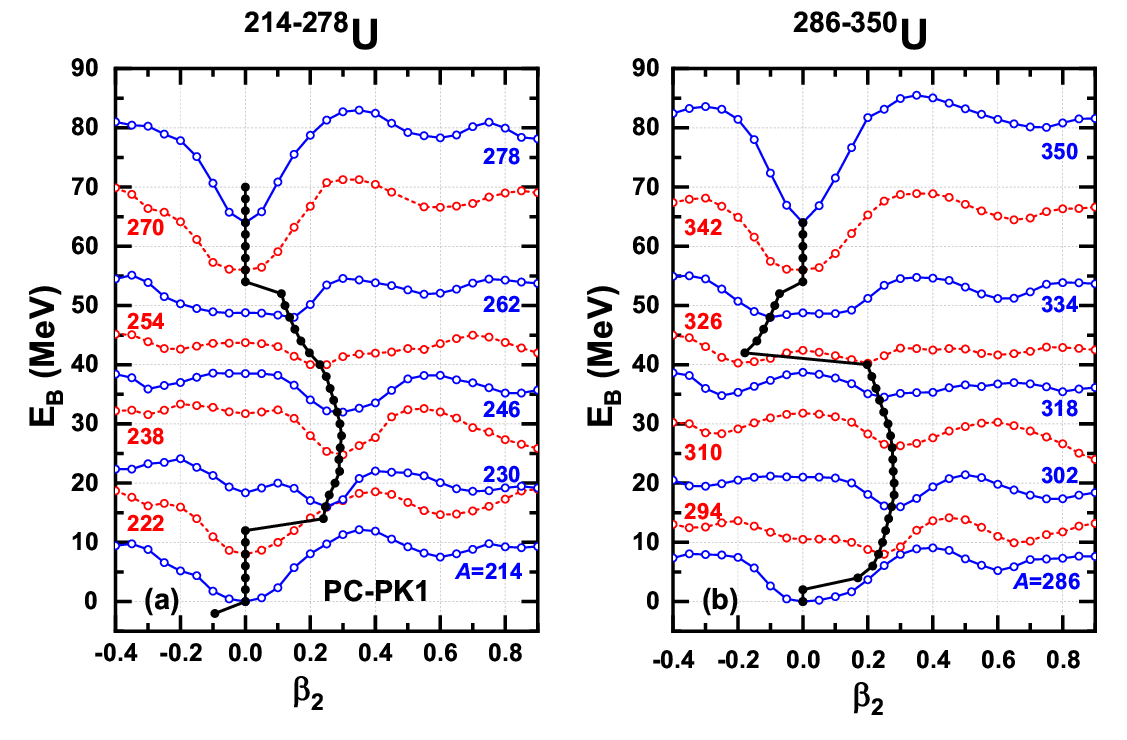}
\caption{(Color online) PECs of $^{214-278}$U~(a) and $^{286-350}$U~(b) denoted by open circles by the constrained DRHBc calculations with PC-PK1. For comparison, the ground state deformations are shown as solid circles obtained by unconstraint calculations. In each panel, the PECs of $^{214}$U and $^{286}$U are renormalized to their ground states, and others are shifted upward one by one by 8~MeV.}
\label{PEC}
\end{figure*}

The numerical details of the present DRHBc calculations for the ground states of uranium isotopes follow those of the DRHBc mass table construction~\cite{ZhangKY2022}. The point-coupling density functional PC-PK1~\cite{Zhao2010} is adopted. The size of the coordinate space is $R_{\rm box} = 20$~fm,
the cutoffs for the energy and angular momentum in the DWS basis are $E_{\rm cut} = 300$ MeV and $J_{\rm max} = 23/2\hbar$, and the Legendre expansion truncation order in Eq.~(\ref{lmax}) is $\lambda_{\rm max}=8$. The pairing strength is $V_0=-325~\rm {MeV~fm}^3$ along with the pairing window of 100 MeV. However, for the calculation of potential energy curves~(PECs), as large quadrupole deformation is involved, a larger $\lambda_{\rm max} = 10$~\cite{Congpan2019} and a larger $J_{\rm max}=31/2 \hbar$ are necessary. As the zero-range pairing interaction is used here in the particle-particle channel, a weaker pairing strength $V_0= -300$ MeV fm$^3$ is adjusted in order to reproduce the ground state energies of uranium isotopes.

Figure~\ref{PEC} shows the PECs obtained by the constrained DRHBc calculations for the uranium isotopes from $A=$ 214 to 350 with an interval of $\Delta A=8$.
The ground state deformations obtained in the unconstraint calculations are also shown for all even-even isotopes
from the proton drip line $^{212}$U to the neutron drip line $^{350}$U predicted by DRHBc in Ref. ~\cite{ZhangKY2022}.
It can be seen that these ground states go exactly through all the global minima on the individual PECs,
which cross-checks the correctness of the constrained calculations. It is noted that the predicted ground state deformations for $^{230,232,234,236,238}$U are in good agreement with the empirical data extracted from the experimental $B(E2)$ values, and they are 0.246, 0.258, 0.276, 0.290, 0.288 while 0.260, 0.264, 0.266, 0.274, 0.274 for empirical data~\cite{Pritychenko2016}. A periodic pattern is exhibited for the shape evolution with neutron number, i.e., for closed shells $N=$126, 184, and 258, the ground states of $^{218,276,350}$U and their neighbors are spherical; away from $N=$126 to the next closure $N=$184, the ground states experience spherical, prolate, and to spherical shape again; and away from $N=$184 to the next closure $N=$258, they experience spherical, prolate, oblate, and then back to spherical shape. It is noted that the prominent shape transition from prolate to oblate occurs from $^{326}$U~($\beta_2=0.198$) to $^{328}$U~($\beta_2=-0.179$). Particularly for the PEC of $^{326}$U, the prolate and oblate minima have very close energies with a difference of 0.05 MeV and a low barrier with a height of 2.18 MeV locating between them, which indicates possible shape coexistence.

\begin{figure}[t!]
\includegraphics[width=0.95\linewidth]{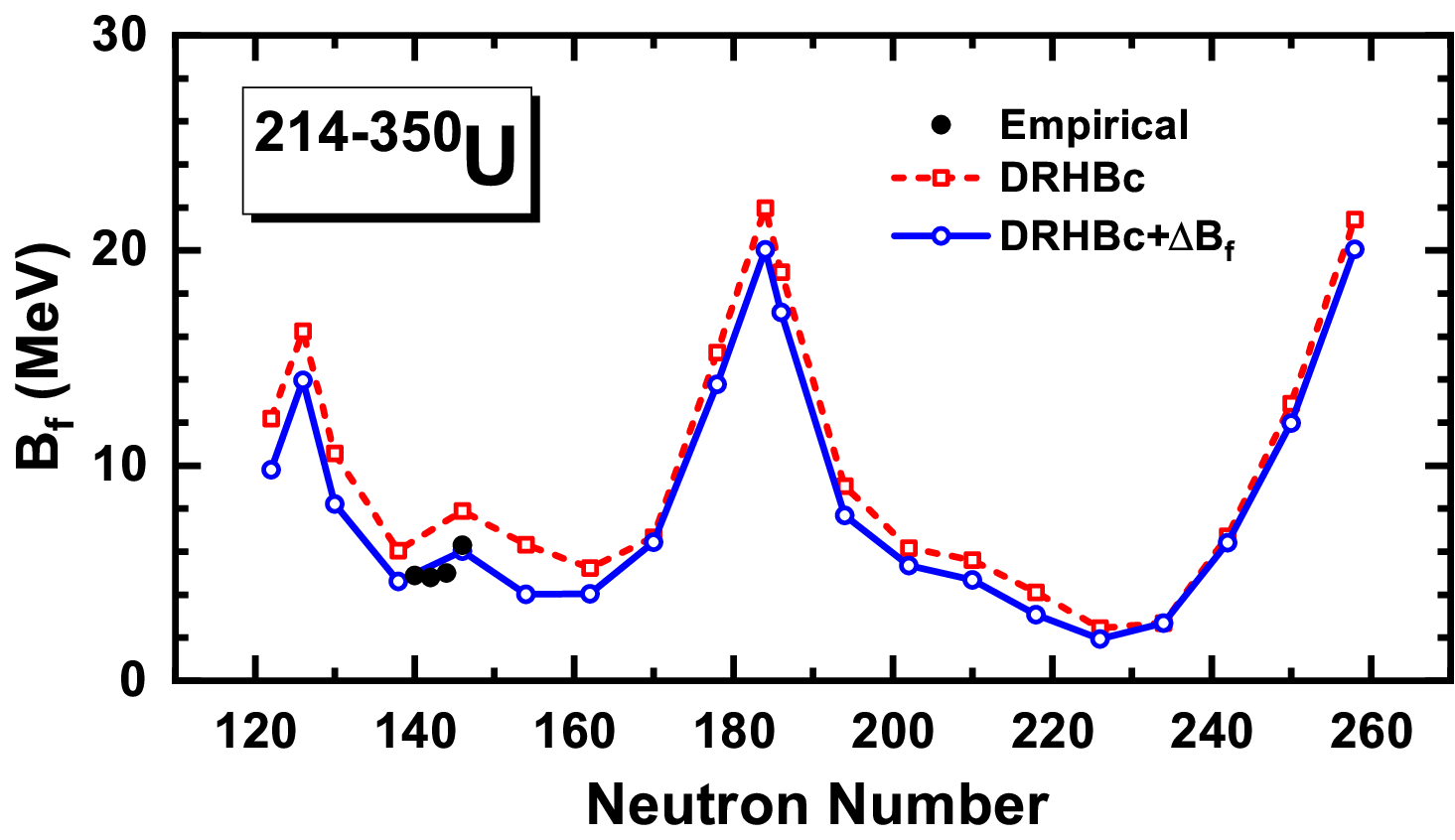}
\caption{(Color online) Inner fission barrier heights $B_{\rm f}$ in the DRHBc calculations for even-even U isotopes, in comparison with the empirical data~\cite{RIPL2009}. The results with the triaxial correction $\Delta B_{\rm f}$ are also shown.}
\label{Bf}
\end{figure}

In Fig.~\ref{Bf},
the inner fission barrier heights $B_{\rm f}$ obtained in the axially deformed case by DRHBc are shown,
which is taken as the energy difference between the ground state and the top of the first barrier.
In analogy to the ground state shape evolutions, the fission barrier height $B_{\rm f}$ also manifests a distinct pattern. The $B_{\rm f}$ shows pronounced peaks ($\approx 20$ MeV) when the neutron numbers are at the shell closures, i.e., $N=$126, 184, and 258 while in the mid-shell, the $B_{\rm f}$ is lowered dramatically with a minimum of 2.44 MeV at $^{318}$U. For comparison, the available empirical fission barriers obtained by fitting experimental fission cross sections~\cite{RIPL2009} are also shown. It can be seen that the fission barrier heights extracted for $^{232,234,236,238}$U from the DRHBc theory are about 2 MeV higher than the empirical values, the main reason of which is believed to be triaxial effect.

\begin{figure*}[t!]
\includegraphics[width=0.8\linewidth]{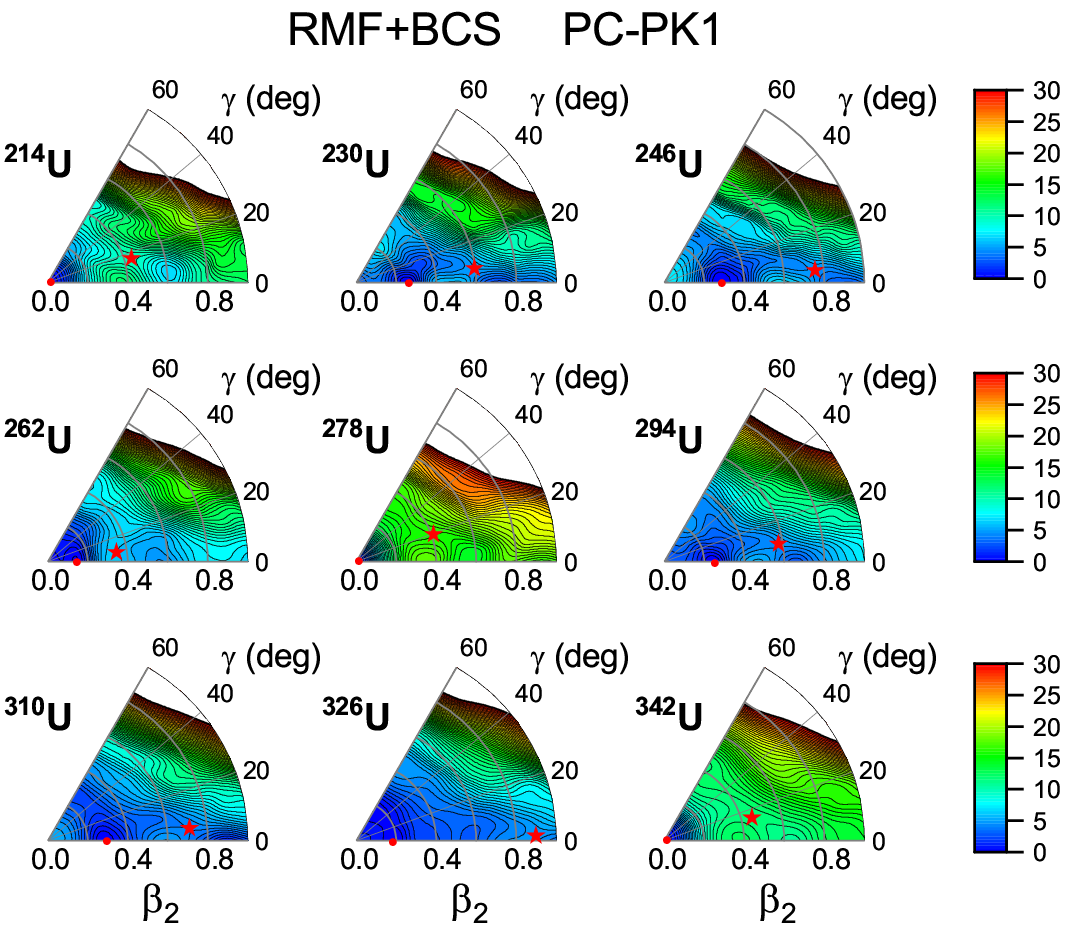}
\caption{(Color online)
PESs in the $(\beta_2,\gamma)$ plane for $^{214-342}$U isotopes with $\Delta A $=16,
calculated by the RMF + BCS model. For each nucleus, the global minimum and saddle point are represented by the red dots and stars, respectively. The energy separation between contour lines is 0.5 MeV.}
\label{UPES}
\end{figure*}

According to the previous studies, the importance of triaxiality to reduce the inner fission barrier height
has been emphasized ~\cite{Moller1970,Randrup1976,Girod1983,Rutz1995,Moller2009,Abusara2010,Schunck2014,Zhou2016}.
In principle, we need explicitly consider the triaxial effect for the study of inner barrier height by taking models such as the triaxial relativistic Hartree-Bogoliubov theory (TRHB)~\cite{TRHBc}. However, we met the difficulty of huge computation. Instead, a simple model, i.e., RMF+BCS in the harmonic oscillator basis, is taken to extract the triaxial effect in this work. To keep consistency with the DRHBc calculations, the same density functional PC-PK1 is taken. The harmonic oscillator basis is $N_{\rm f}=16$ while the parameters of the separable pairing force $G$ = 728 MeV fm$^3$ and $a$ = 0.644 fm are taken from Ref.~\cite{YTian2009}.

In Fig.~\ref{UPES}, the potential energy surfaces (PESs) in the ($\beta_2$, $\gamma$) plane obtained by the constrained RMF + BCS calculations are shown for $^{214-342}$U with $\Delta A=16$. It can be found that the ground states of the U isotopes denoted by red dots are all spherical or axially deformed, which verifies the appropriateness of the ground state deformation obtained in the DRHBc calculations presented in Fig.~\ref{PEC}. Particularly for $^{326}$U where the prolate and oblate minima are very close in energy in Fig.~\ref{PEC}, the triaxial RMF + BCS calculations also predict a prolate minimum at $\beta_2 \sim 0.20$ and an oblate minimum at $\beta_2 \sim -0.20$ with a tiny energy difference of 50 keV. This means the prediction of shape coexistence in $^{326}$U still holds with the triaxial degree of freedom taken into account. In Fig.~\ref{UPES}, we also mark out the position of the saddle point for each nucleus by red stars. It can be seen that the $\beta_2$ values of these saddle points locate between 0.4 and 0.8, while the corresponding $\gamma$ values locate between 0$^\circ$ and 20$^\circ$. It is clear that the saddle point is lower than the peak with axial symmetry, and the consideration of the triaxial degree of freedom does reduce the height of the inner fission barrier for the uranium isotopes.

\begin{figure}[t!]
\centering
\includegraphics[width=0.95\linewidth]{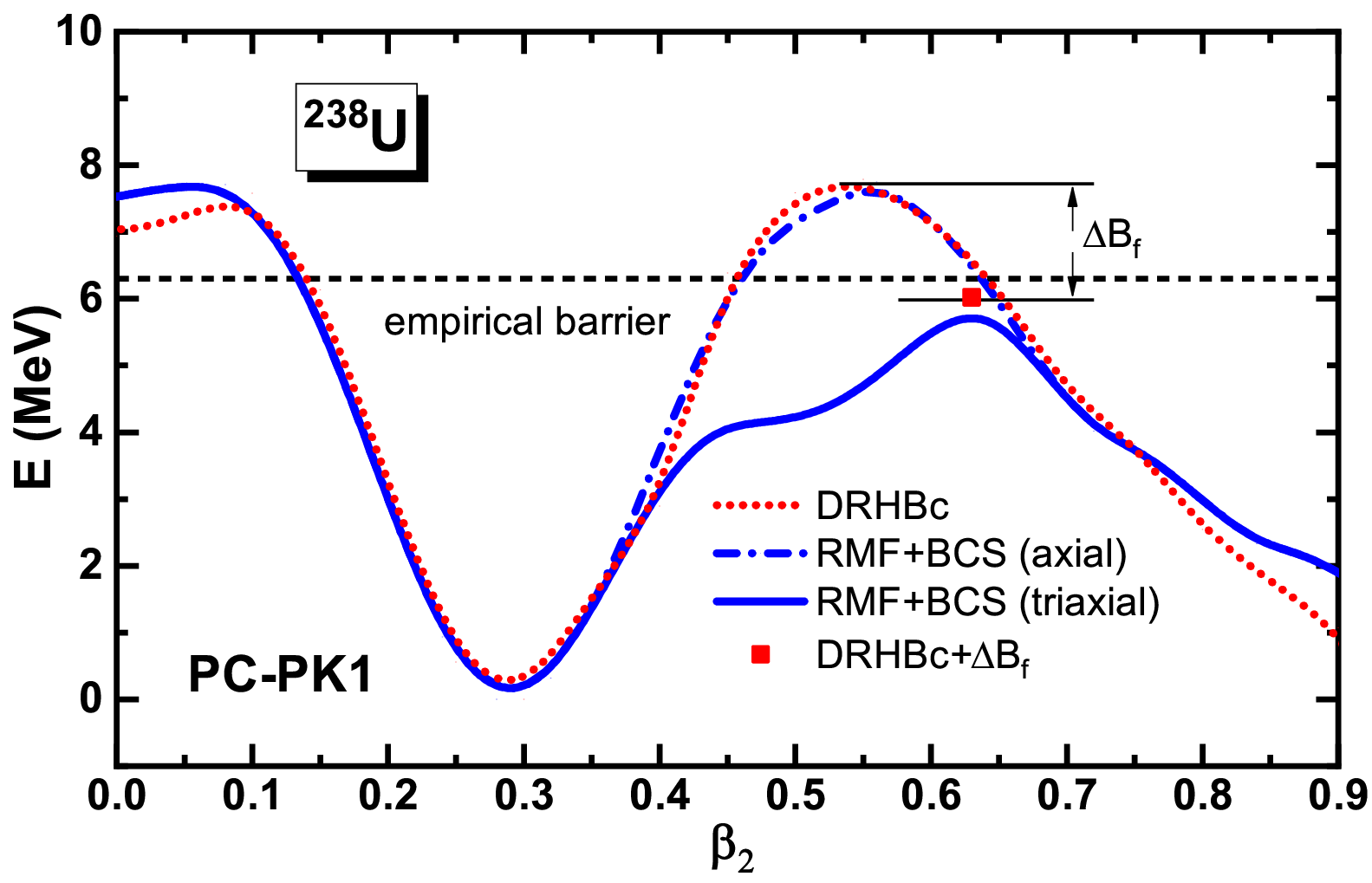}
\caption{(Color online) PECs of $^{238}$U by DRHBc~(red dotted line), RMF+BCS under axial symmetry~(blue dash-dotted line), and triaxial RMF+BCS~(blue solid line) calculations. The triaxial correction is determined by $\Delta B_{\rm f}=B_{\rm f}^{\rm BCS}({\rm triaxial})-B_{\rm f}^{\rm BCS}({\rm axial})$. The black dashed line denotes the empirical barrier.}
\label{U238Bf}
\end{figure}

Based on the RMF+BCS calculations, the reduction of the fission barrier by triaxial effect is defined as
$\Delta B_{\rm f}=B_{\rm f}^{\rm BCS}({\rm triaxial})-B_{\rm f}^{\rm BCS}({\rm axial})$ with $B_{\rm f}^{\rm BCS}({\rm triaxial})$ being the height of the fission barrier in the PES determined by the triaxial calculations and $B_{\rm f}^{\rm BCS}({\rm axial})$ the corresponding height under the restriction of axial symmetry. Taking $^{238}$U as an example, Fig.~\ref{U238Bf} illustrates the PECs obtained by the DRHBc and triaxial RMF + BCS calculations. For the latter, both the PEC along the fission path (blue solid line) and the one under the axial symmetry (blue dash-dotted line) have been plotted.
In general, the PEC by the DRHBc is in good agreement with the one by the RMF + BCS under the axial symmetry except for the perceptible differences in the regions of $\beta_2\le 0.10$ and $\beta_2\ge 0.75$. Obvious triaxial effects are found in the range of $0.40 \le\beta_2\le 0.65$ which significantly reduce the fission barrier by an amount of $-\Delta B_{\rm f}=1.87$ MeV. After superimposing the $\Delta B_{\rm f}$ to the fission barrier height obtained by the DRHBc calculation, the corrected barrier height denoted by the red square becomes very close to the empirical value.

\begin{figure}[t!]
\includegraphics[width=0.95\linewidth]{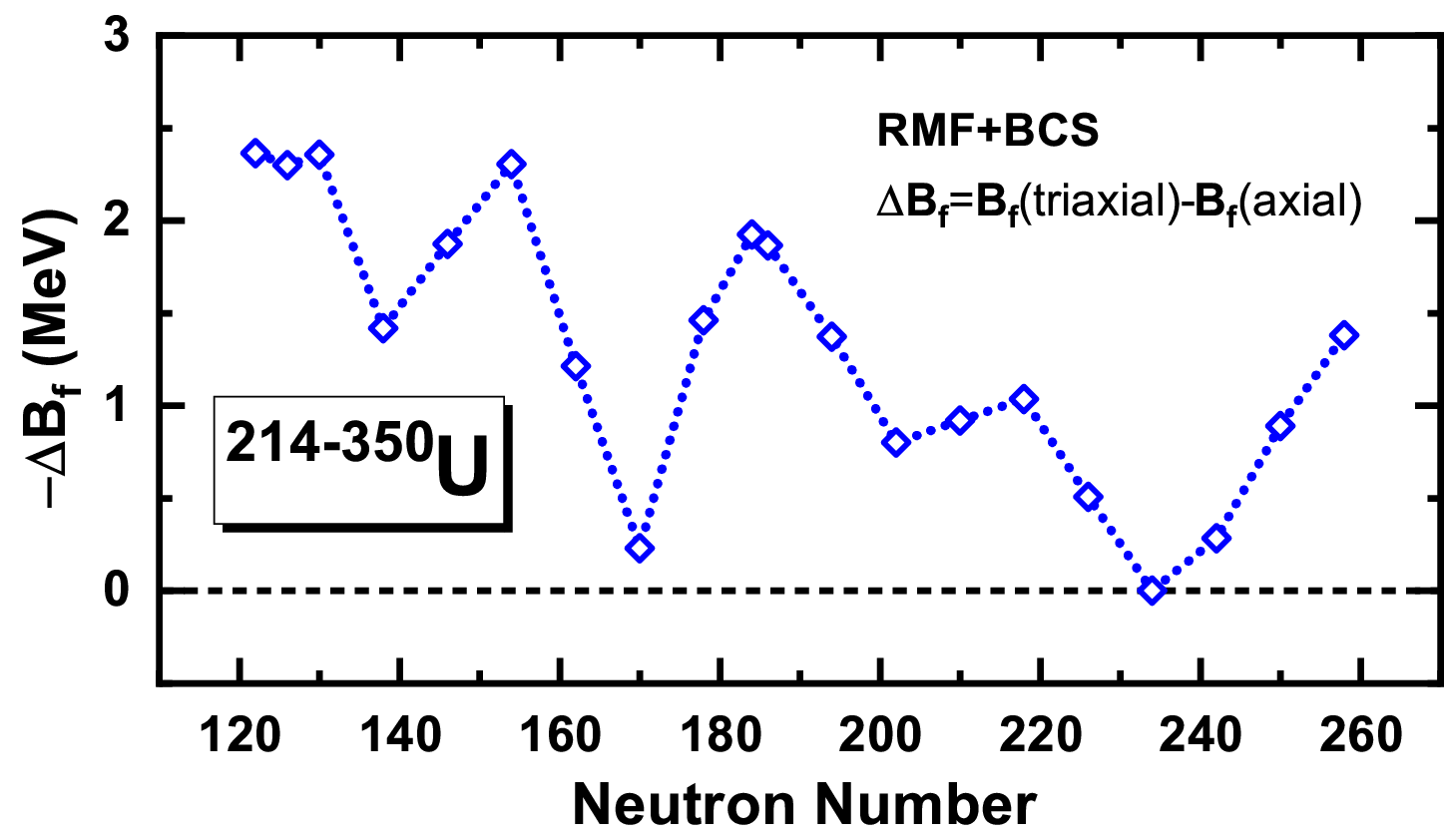}
\caption{(Color online) Triaxial correction $-\Delta B_{\rm f}$ to
the inner fission barrier by the RMF+BCS model as a function of the neutron number.}
\label{DBf}
\end{figure}

\begin{figure}[t!]
\includegraphics[width=0.95\linewidth]{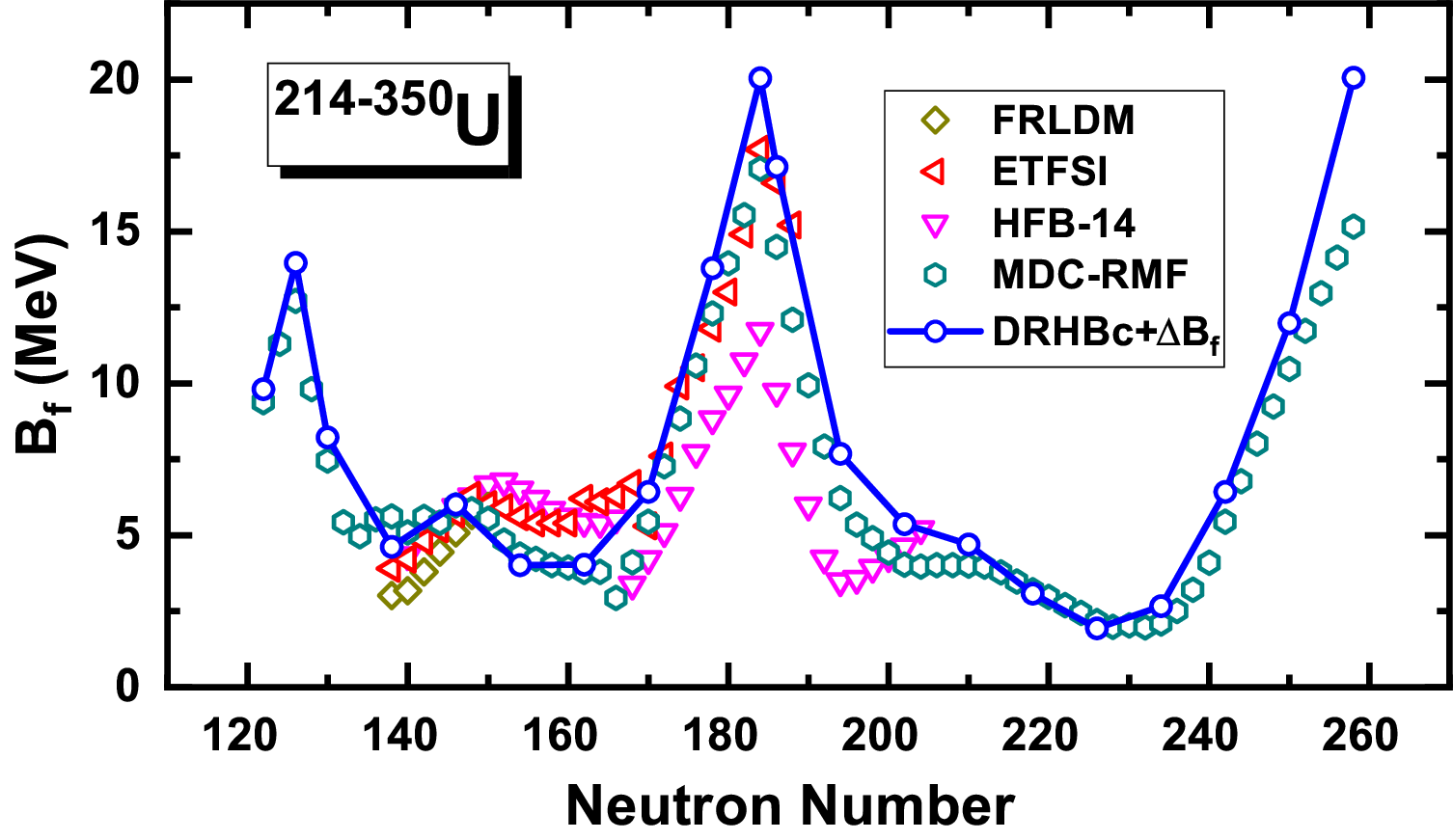}
\caption{(Color online) Inner fission barrier heights for even-even U isotopes in the DRHBc calculations with the triaxial correction, in comparison with the results by FRLDM~\cite{Moller2009}, ETFSI~\cite{MAMDOUH2001}, HFB-14~\cite{Goriely2009} and MDC-RMF~\cite{Deng2023} models. }
\label{TheBf}
\end{figure}

Taking the prescription shown in Fig.~\ref{U238Bf}, the triaxial corrections $\Delta B_{\rm f}$ in the fission barrier have been estimated for the $^{214-350}$U isotopes plotted as a function of neutron number in Fig.~\ref{DBf}. Obvious different triaxiality is shown for uranium isotopes, e.g., the correction $-\Delta B_{\rm f}$ can be as large as 2.5 MeV for $^{214}$U while almost zero for $^{326}$U. For $^{214}$U, as seen in Fig.~\ref{UPES}, the saddle point denoted by the red star is with a large triaxial deformation, which means the static fission path in the $(\beta_2, \gamma)$ plane is far away from the axial symmetric path, leading to a remarkable triaxial effect. However, for $^{326}$U, the saddle point and the fission path are with the axial symmetry, thus the triaxial effect is zero.
By superimposing $\Delta B_{\rm f}$ to the barrier height obtained by the DRHBc theory, in Fig.~\ref{Bf}, good agreement with the empirical fission barriers of $^{232,234,236,238}$U is obtained. Besides, the triaxial effect does not change the global trend of the barrier height.

Finally, in Fig.~\ref{TheBf}, inner fission barriers by the DRHBc theory with the triaxial correction $\Delta B_{\rm f}$ are compared with the results by FRLDM~\cite{Moller2009}, ETFSI~\cite{MAMDOUH2001}, HFB-14~\cite{Goriely2009} and MDC-RMF~\cite{Deng2023} models. For isotopes with $ 138 \leqslant N \leqslant 194$, the evolutions of the fission barrier heights given by these theories are analogous to each other. For the peak at $N=184$, the DRHBc gives the largest barrier height, the HFB-14 gives the lowest one, and the ETFSI and MDC-RMF are in between. Drastic different evolutions of the barrier heights appear around $N \geqslant 194 $, i.e., results by the HFB-14~\cite{Goriely2009} increase with the neutron number $N$ while those by the MDC-RMF~\cite{Deng2023} and DRHBc decrease with $N$. However, from the proton drip line to the neutron drip line, the DRHBc and MDC-RMF results show remarkable consistency. It is further noted that a flat valley is shown around the neutron-rich nucleus $^{318}$U with the barrier height as small as 1.9 MeV, indicating a large probability in spontaneous or induced fission which may play an important role in the astrophysical $r$-process nucleosynthesis~\cite{Kajino2019}.

\section{Summary and perspective}
\label{section:Summary}
In this work, we investigate the shape evolution and the inner fission barriers of the even-even uranium isotopes from the proton to the neutron drip line with the deformed relativistic Hartree-Bogoliubov theory in continuum. The obtained ground state deformations of $^{230,232,234,236}$U are in good agreements with the empirical data. A periodic evolution is shown for the ground state shape
with the neutron number: it evolves from spherical at the closed shell $N = 126$ to prolate at the mid-shell, then back to spherical near the next closed shell $N = 184$, after that it further undergoes spherical to prolate to oblate transitions, and finally becomes spherical again near the next possible closed shell $N = 258$. From the PECs obtained by the DRHBc calculations, the inner fission barriers of uranium isotopes have been extracted. The triaxial correction to the inner fission barrier is further evaluated using the triaxial RMF + BCS method. With the triaxial correction included, good consistency with the available empirical inner barrier heights is found. Besides, the evolutionary trend from the proton to neutron drip line is in accord with the results by the MDC-RMF theory. Note that a flat valley in the fission barrier height is predicted around the neutron-rich nucleus $^{318}$U, indicating that a possible fission intends to happen easily which may play a key role in the astrophysical $r$-process nucleosynthesis.

\section*{Acknowledgments}
The authors thank Dr. Xiangquan Deng for providing us the inner fission barriers of uranium nuclei with the MDC-RMF calculations. Helpful discussions with members of the DRHBc Mass Table Collaboration are highly appreciated. This work was partly supported by the Natural Science Foundation of Henan Province (Grant No.~242300421156,~202300410480), the National Natural Science Foundation of China (Grant No.~12141501,~U2032141,~11935003), the State Key Laboratory of Nuclear Physics and Technology, Peking University (Grant No.~NPT2023ZX03), the Super Computing Center of Beijing Normal University,
and High-performance Computing Platform of Peking University.

\vspace{-1mm}
\centerline{\rule{80mm}{0.1pt}}
\vspace{2mm}



\clearpage

\end{document}